\documentclass[sigconf]{acmart}

\usepackage{xspace}
\usepackage{enumitem}

\usepackage{tcolorbox}
\definecolor{tagbordercolor}{HTML}{B0BEC5}
\definecolor{tagbgcolor}{HTML}{ECEFF1}
\definecolor{hotpink}{RGB}{255, 83, 115}

\newtcbox{\captag}{nobeforeafter, colframe=tagbordercolor,
colback=tagbgcolor, boxrule=0.5pt, arc=1pt,
 boxsep=0pt,left=2pt,right=2pt,top=1.5pt,bottom=2pt,tcbox raise base}

\newcommand{\argo}{\textsc{Argo Lite}\xspace}

\definecolor{linkColor}{RGB}{6,125,233}

\AtBeginDocument{%
  \providecommand\BibTeX{{%
    \normalfont B\kern-0.5em{\scshape i\kern-0.25em b}\kern-0.8em\TeX}}}

\copyrightyear{2020}
\acmYear{2020}
\setcopyright{acmcopyright}\acmConference[CIKM '20]{Proceedings of the 29th ACM International Conference on Information and Knowledge Management}{October 19--23, 2020}{Virtual Event, Ireland}
\acmBooktitle{Proceedings of the 29th ACM International Conference on Information and Knowledge Management (CIKM '20), October 19--23, 2020, Virtual Event, Ireland}
\acmPrice{15.00}
\acmDOI{10.1145/3340531.3412877}
\acmISBN{978-1-4503-6859-9/20/10}

\begin{document}
\fancyhead{}
\title{Argo Lite: Open-Source Interactive Graph Exploration and Visualization in Browsers}

\author{Siwei Li, Zhiyan Zhou, Anish Upadhayay, Omar Shaikh, Scott Freitas, Haekyu Park, Zijie J. Wang, Susanta Routray, Matthew Hull, Duen Horng Chau}
\affiliation{%
  \institution{Georgia Institute of Technology}
  \streetaddress{North Ave NW}
  \city{Atlanta}
  \state{Georgia}
  \country{United States}}
\email{{robertsiweili, zzhou406, aupadhayay3, oshaikh, safreita, haekyu, jayw, sroutray3, matthewhull, polo}@gatech.edu}

\renewcommand{\shortauthors}{Li, et al.}

\begin{abstract}
Graph data have become increasingly common. Visualizing them helps people better understand relations among entities.
Unfortunately, existing graph visualization tools are primarily designed for single-person desktop use,
offering limited support for interactive web-based exploration and online collaborative analysis.
To address these issues, we have developed
\argo{}, a new in-browser interactive graph exploration and visualization tool.
\argo{} enables users to publish and share interactive graph visualizations as URLs and embedded web widgets.
Users can explore graphs incrementally by adding more related nodes, such as highly cited papers cited by or citing a paper of interest in a citation network. 
\argo{} works across devices and platforms, leveraging WebGL for high-performance rendering.
\argo{} has been used by over 1,000 students at Georgia Tech's \textit{Data and Visual Analytics} class.
\argo{} may serve as a valuable open-source tool for advancing multiple CIKM research areas, from \textit{data presentation}, to \textit{interfaces for information systems} and more.
\end{abstract}



\begin{teaserfigure}
  \includegraphics[width=\textwidth]{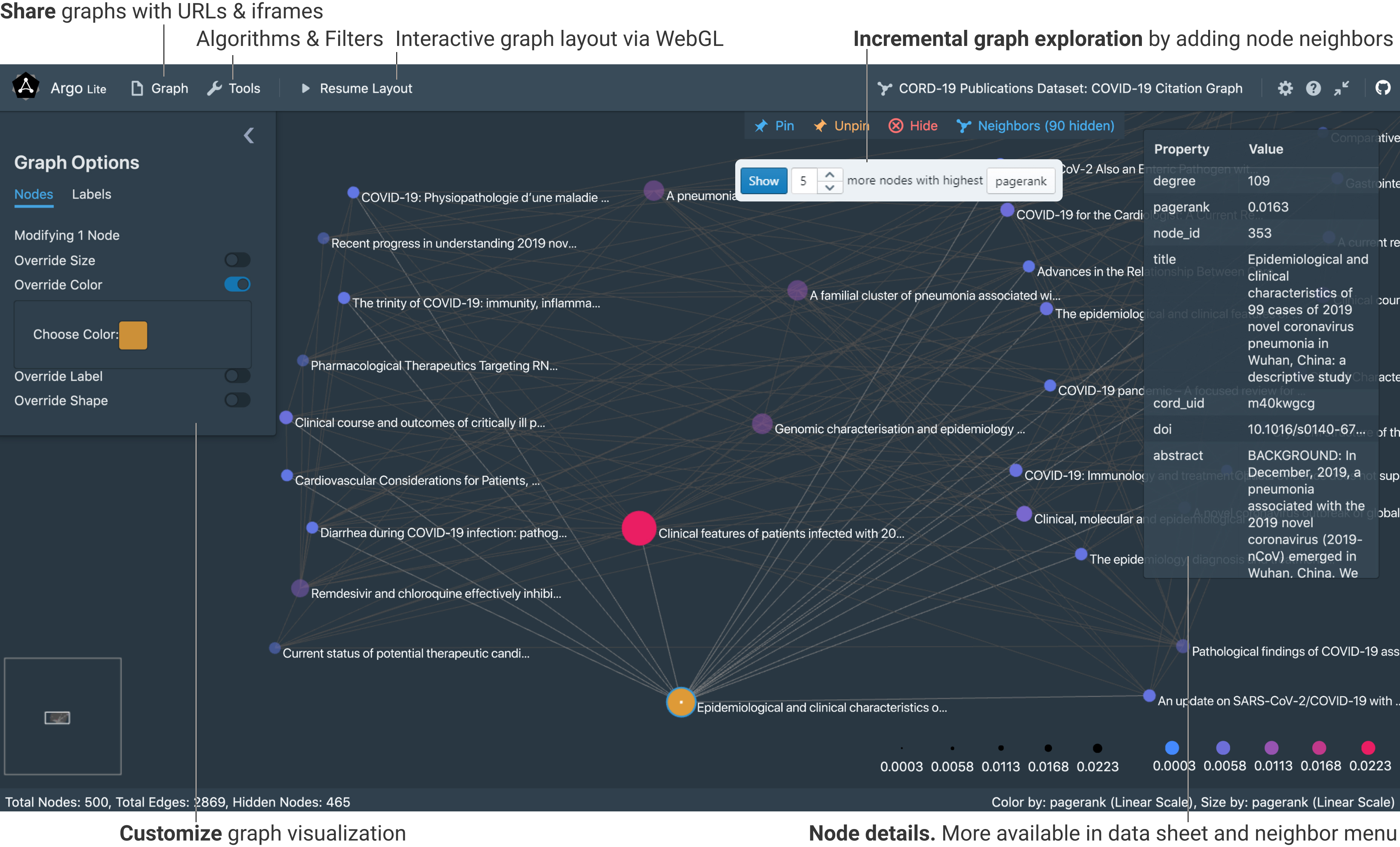}
  \caption{
  \argo{} visualizing a citation network of recent COVID-19 publications \cite{Wang2020CORD19TC}.
  \argo{} users can explore graphs incrementally by adding more related papers (e.g., highly cited papers cited by or citing a paper of interest) to the visualization \cite{chau2011apolo}.
  Using WebGL for high-performance cross-platform graph rendering \cite{Carina}, \argo{} runs in all modern web browsers without requiring any installation.
  }
  \Description{A screenshot of Argo Lite visualizing a citation graph of COVID-19 publications.}
  \label{fig:teaser}
\end{teaserfigure}

\maketitle

\section{Introduction}

As network data become increasingly common, many software tools have been developed to help researchers better understand them and uncover insights. Tools such as Gephi \cite{bastian2009gephi} and Cytoscape \cite{shannon2003cytoscape} use visualizations to help reveal the relationship among entities. However, existing graph visualization and exploration tools face a variety of challenges that limit their availability, interactivity and collaboration productivity.

\subsection{Challenges of Existing Systems}

\begin{enumerate}[label=C\arabic*.,itemsep=2mm, topsep=2mm, parsep=1mm, leftmargin=6mm, label=\textbf{C\arabic*.}, ref=\textbf{C\arabic*}]

\item \label{item:c1} %
\captag{\sffamily \textsc{\small{Availability}}}
\textbf{Device, platform and OS constraints.}
Existing graph visualization tools are typically built for desktop use only \cite{bastian2009gephi,shannon2003cytoscape,chau2011apolo},
making them difficult to access via the Web or on mobile devices, such as tablets and Chromebooks. They require heavyweight installation processes and are only available on certain operating systems.

\item \label{item:c2} %
\captag{\sffamily \textsc{\small{Interactivity}}}
\textbf{Lack of fast interactive web visualization.}
While research has shown that interactive  visualizations can help users more easily and rapidly make sense of graph data \cite{chau2011apolo},
it has been challenging to support interactive explorable graph visualizations on the Web. 
Currently, people who wish to publish graph visualizations in digital formats---such as in online articles, blog posts, or via Jupyter Notebooks---are often limited to using static images.
To provide interactivity, researchers often resort to writing custom JavaScript-based visualizations using libraries such as D3 \cite{d3}, which requires complex engineering efforts; furthermore, the resulting visualizations may not scale beyond 
a few hundred nodes \cite{Carina}.

\item \label{item:c3} %
\captag{\sffamily \textsc{\small{Collaboration Productivity}}}
\textbf{Difficulty to share and collaborate.}
Graph visualization tools have traditionally been built as single-user software \cite{pienta2015scalable}.
They lack modern collaboration features, such as saving progress ``in the cloud,'' or sharing via URL links and as embedded web widgets.
Instead,  traditional tools rely on saving files onto local file systems; users would need to manually pass these files back and forth among collaborators.
Furthermore, people who wish to publish their data and visualizations need to figure out where and how to host their files; and the recipients need to download all the required files and software before they can see the visualizations.
Such user experience is a departure from what many people are now accustomed to when using   
modern applications like Google Docs, where users can share their documents through URLs and no installation or file download is required for viewing.

\end{enumerate}

\subsection{Argo Lite Novel Contributions}

We introduce \argo{} (\autoref{fig:teaser}), an open-source in-browser interactive graph visualization and exploration system developed using the latest web technologies. Our main contributions are:

\begin{itemize}[label=C\arabic*.,itemsep=2mm, topsep=2mm, parsep=1mm, leftmargin=4mm,]

\item[\textbf{1.}]
\textbf{Easy-to-access platform-agnostic web application.}
\argo{} is a web application that runs on all modern browsers (Chrome, Firefox, Safari, Edge, etc.). 
Users can access \argo{} on a wide range of devices, from desktop computers, Chromebooks, to tablets, regardless of the operating system (\ref{item:c1}).
\argo{} is freely accessible at \textcolor{linkColor}{\url{https://poloclub.github.io/argo-graph-lite}}. 
\argo{}'s code repository and documentation are available at \textcolor{linkColor}{\url{https://github.com/poloclub/argo-graph-lite}}.

\item[\textbf{2.}]
\textbf{Fast graph rendering via WebGL.}
\argo{} takes advantage of the WebGL technology to drastically improve graph rendering performance in web browsers (\ref{item:c2}).
Prior systems such as Cytoscape Web \cite{lopes2010cytoscapeweb} use Adobe Flash, now a deprecated technology for interactive content for the Web.
Most web-based visualization libraries today use D3 \cite{d3} and render graph using either \textit{Scalable Vector Graphics}~(SVG) or HTML Canvas Elements. The rendering performance, measured by frame rate per second (FPS), starts to deteriorate significantly at a low hundreds of nodes and edges \cite{Carina}. 
\argo{} uses WebGL to render graphs, which leverages hardware acceleration and is supported by all modern web browsers.
\argo{} smoothly renders hundreds of nodes and edges with interactive force-directed layout on commodity devices (e.g., laptops, tablets).

\item[\textbf{3.}]
\textbf{Sharing interactive graph visualizations via links and embedded widgets.}
\argo{} allows users to share interactive graph visualizations as a URL or embedded web widget based on iframe. Users can publish their interactive graph visualizations on their web articles, blog posts or even interactive notebooks (such as a Jupyter Notebook). Readers can explore the same interactive visualization directly through the embedded web widget, without the need to install any software or to download any file (\ref{item:c1}, \ref{item:c3}). Researchers can easily generate a link\footnote{\textcolor{linkColor}{\url{https://poloclub.github.io/argo-graph-lite/\#6c0d8aaa-5320-4c81-9618-11ea5e0524f4}}} for their new visualization or exploration ``snapshot.'' 
Sharing the link
with collaborators would allow them to access, modify, and continue their explorations on this graph data. 
This greatly facilitates collaboration within teams and among graph and network researchers (\ref{item:c3}).

\item[\textbf{4.}]
\textbf{Broadening impact of graph analysis.}
\argo{} provides an easy-to-use experience that helps  make the study of graph data more impactful. It is designed not just for experts, but brings interactive graph visualization to a wider audience.

\argo{} has been successfully used by over 1,000 students at Georgia Tech's \textit{Data and Visual Analytics} class, where students have used the force-directed layout, sizing, coloring and labeling features to produce different visualizations and then shared them as URLs using the snapshot sharing feature.

To help researchers get started, we have provided several sample graph datasets,
including the real-world citation graph of recent
COVID-19 publications (\autoref{fig:teaser}) derived from the CORD-19 Open Research Dataset \cite{Wang2020CORD19TC}.
Because the dataset includes papers from many disciplines (e.g., biomedical, mathematical modeling, machine learning), not all papers are strongly related to every user's research interest.
To support more personalized use of the dataset, \argo{} allows users to easily hide papers less relevant to their interests, and enables them to incrementally add more related papers as they expand their exploration (e.g., add highly cited papers of a paper of interest).
Users can also easily share the exploration subgraph results as a graph snapshot URL.
We believe the impact of \argo{} will grow all the more, and empower more users to benefit from interactive graph visualization.

\end{itemize}

\section{Argo Lite: Feature Highlight}
\subsection{In-browser and Cross-platform}

\argo{} is a web application that runs on modern browsers and operating systems.
It runs on popular web browsers such as Chrome, Firefox, Safari, and Edge, which are available on Windows, Mac OSX, Linux, Android and iOS devices. Since it does not require OS-level installation, users can also run \argo{} on tablets and Chromebooks. Argo Lite has complete support for both mouse and touchscreen interactions, with optional keyboard shortcuts.
\subsection{Importing Graphs}

\begin{figure}[tb]
  \centering
  \includegraphics[width=0.8\linewidth]{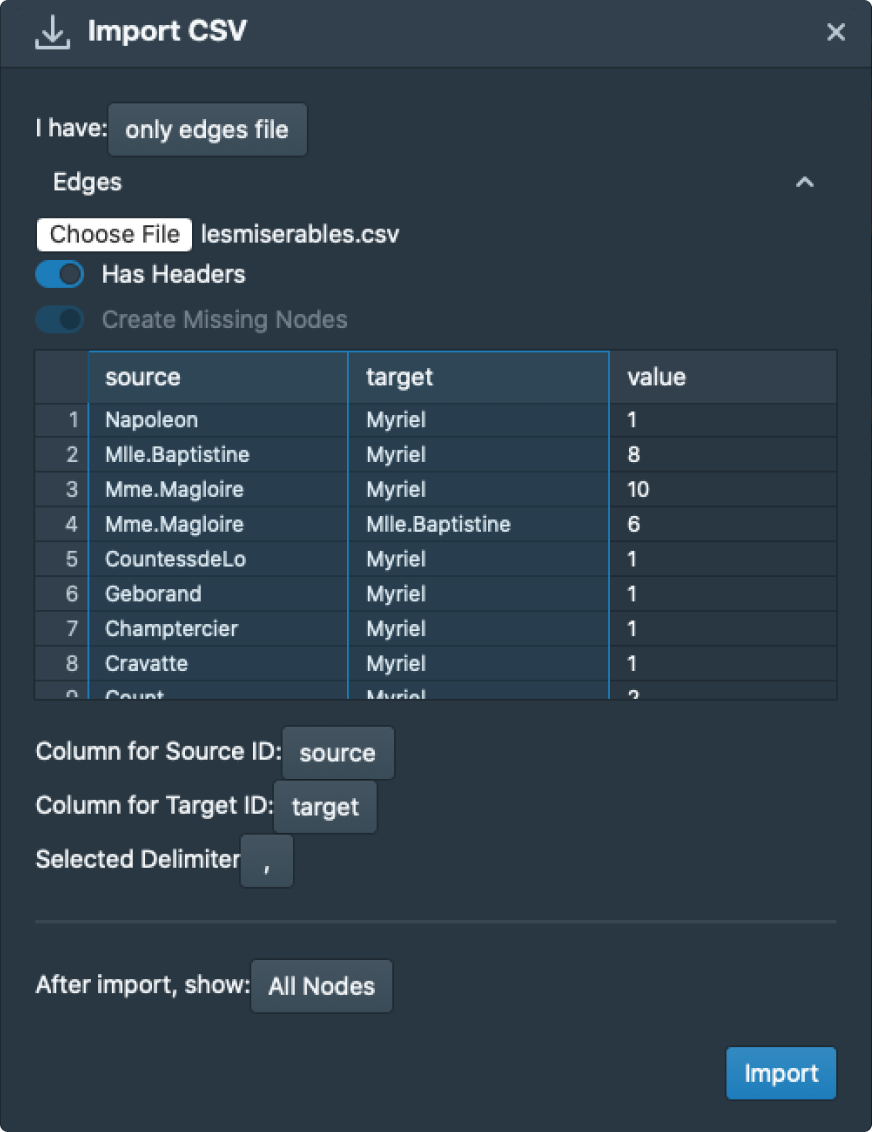}
  \caption{\argo{} helps user import graph from CSV/TSV and GEXF formats. 
  Users can upload and preview the contents of CSV/TSV files in tabular format, select columns to import (e.g., as node IDs) and set additional import options, making it easy to work on graph data from different sources.}
  \Description{Figure showing Argo Lite's import screen.}
  \label{fig:import}
\end{figure}

\argo{} allows user to import graph data from CSV (comma-separated values) and TSV (tab-separated values) files,  as well as the popular GEXF format \cite{bastian2009gephi}. It provides a convenient dialog for users to specify import options and preview the imported data in a tabular format (Figure ~\ref{fig:import}). Users only need to import the data once. From then on, they can work on the data, save graph visualization results and share them as \argo{} \textit{graph snapshots} (Section \ref{subsection:snapshot}) without the need to re-import the data.
\argo{} provides the option for displaying the whole graph after the import, which is appropriate for smaller graphs. For larger graphs, \argo{} provides an option for showing the subgraph with the highest PageRank \cite{page1999pagerank} scores; this serves as a good starting point, as the visual complexity created by the many overlapping edges may overwhelm users \cite{chau2011apolo}.

\begin{figure}[t]
  \centering
  \includegraphics[width=0.9\linewidth]{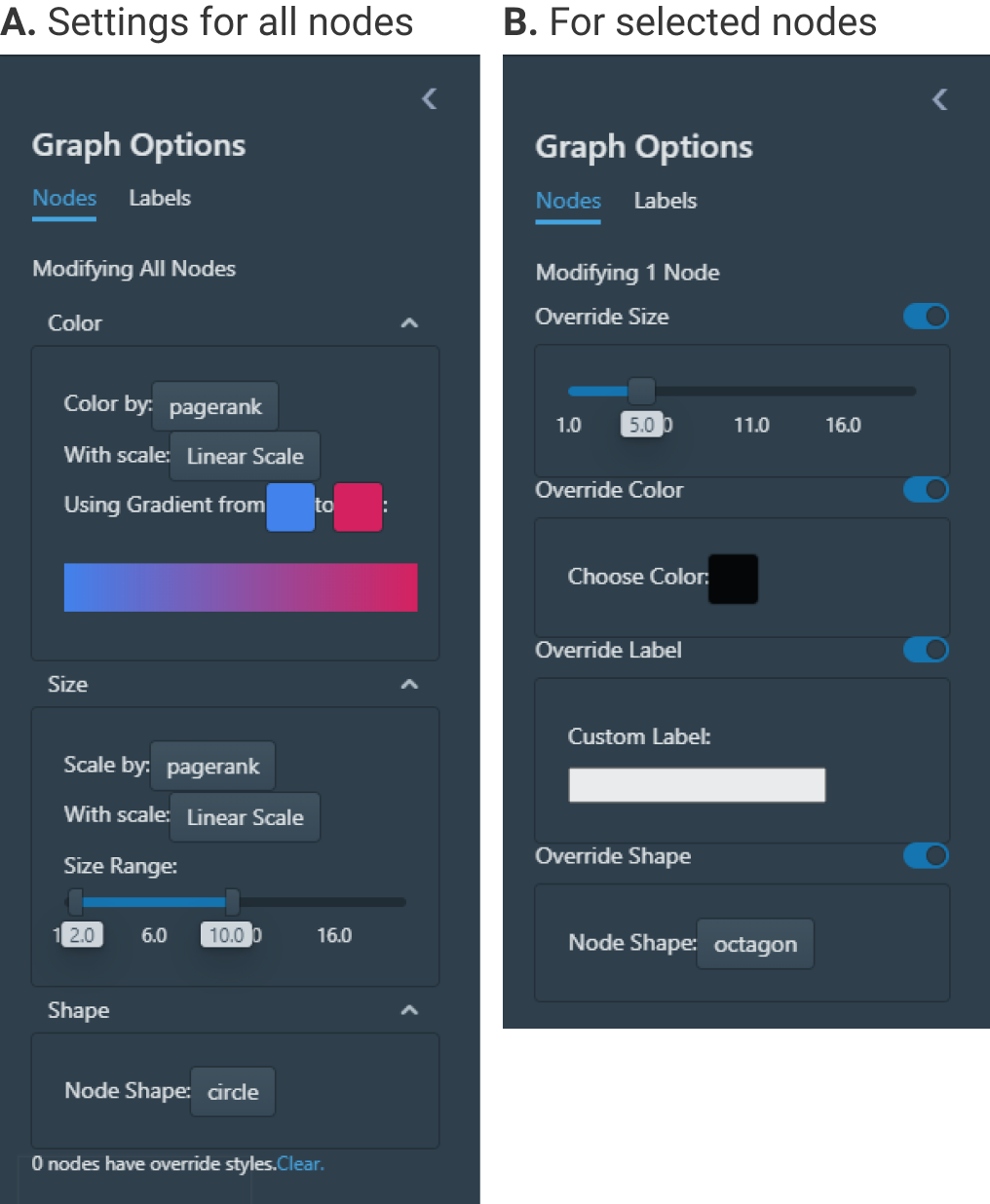}
  \caption{Users have full control over node colors, sizes, shapes and labels. Users can use the global settings to color and size all the nodes by their attributes such as degree and PageRank, or use individual settings to override the style of a node. The diverse options help users more easily create meaningful and informative visualizations.}
  \Description{Figure of Argo Lite's Graph Options panel side by side.}
  \label{fig:visualization}
\end{figure}

\subsection{Visualization}
\label{subsection:vis}

\argo{} provides a variety of built-in helpful features for graph visualization, including force-directed graph layout \cite{fruchterman1991graph}, sizing and coloring by PageRank \cite{page1999pagerank}, degree and other user-specified properties (Figure ~\ref{fig:visualization}A). Users can also customize the visualization of any specific nodes they choose by overriding the nodes' size, color and labels (Figure ~\ref{fig:visualization}B).

\subsection{Incremental Exploration}
\label{subsection:exploration}

\begin{figure}[tb]
  \centering
  \includegraphics[width=0.75\linewidth]{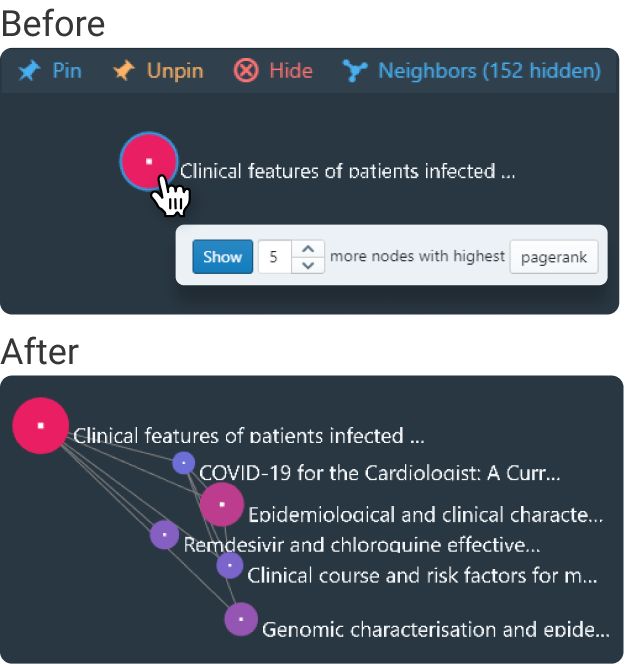}
  \caption{The Neighbor Menu helps user to identify important neighboring nodes of any node selected by the user, and incrementally add them to the visualization. Users can choose to add a set number of neighboring nodes with top PageRank or degree values to quickly navigate the graph, or they can browse the sortable neighbor table to individually pick nodes of interest.}
  \Description{Figure showing Argo Lite's Incremental Exploration workflow.}
  \label{fig:incremental}
\end{figure}

For larger graphs with more than a few thousands of nodes and edges, visualizing the full graph all at once can result in a massive ``hair ball'' with numerous edge crossings that would visually overwhelm the users and make it hard for them to begin their analysis~\cite{chau2011apolo}.
To reduce such visual complexity,  
\argo{} allows users to incrementally explore the graph by starting with a small number of high-PageRank or high-degree nodes \cite{chau2011apolo}.
Users are able to use the \textit{Neighbor} menu or right-click menu to incrementally add neighbor nodes to the graph.
The \textit{Neighbor} menu helps users to identify and add important neighbor nodes, either by PageRank scores or other node attributes (Figure ~\ref{fig:incremental}).

In addition to the \textit{Neighbor} menu, \argo{} also provides a \textit{Data Sheet} (via \textit{Tools} menu) that displays all the nodes in the imported data, giving users the options to inspect and sort them by node attributes, and to start their exploration from specific nodes of interest.

\subsection{Graph Statistics and Algorithms}
\argo{} not only computes PageRank scores \cite{page1999pagerank} and degrees for visualization, but also provides easy access to many other common graph algorithms that help summarize graphs \cite{leskovec2016snap}, including graph density, graph diameter, clustering coefficient and connected components (all available through the \textit{Tools} menu).

\subsection{Saving graph visualization as snapshots}
\label{subsection:snapshot}
Users can save their work as \textit{graph snapshots}, which are JSON documents that store the  full original graph data (e.g., nodes, edges, attributes) along with the \argo{} visualization and exploration state (e.g., node visibility, positions, colors). 
Users can save such \textit{graph snapshots} as files locally, or use the sharing function to save the snapshot ``on the cloud'' which will be described next.

\subsection{Saving \& sharing graph snapshots as URLs}

\begin{figure}[b]
  \centering
  \includegraphics[width=\linewidth]{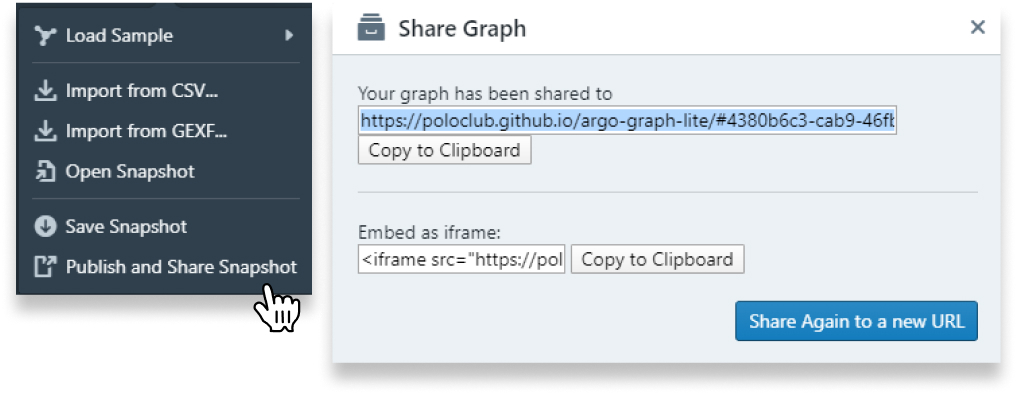}
  \caption{\argo{} allows user to save and share graph snapshots via URL links,  simplifying collaboration.}
  \Description{Figure showing Argo Lite's sharing dialog.}
  \label{fig:share}
\end{figure}

When using a web application, it is useful for users to store their data ``on the cloud,'' and to share them with collaborators as URL links, eliminating the need to download or upload files from local file systems, similar to what many users may be accustomed to when using modern collaboration tools like Google Docs.
\argo{} supports such convenient usage and sharing experiences.
It allows users to save \textit{graph snapshots} (as described in Section \ref{subsection:snapshot}) containing both graph data and visualization states on a backend server and share them as URLs (\autoref{fig:share}). 
Users can continue their exploration from a graph snapshot URL, without having to save the snapshot on local disk.
We have set up a public server that everyone can use for free for public data.
Users can also easily set up their own preferred backend server for sharing sensitive data and exploration results.
As part of \argo{}'s documentation, we have provided a guide for users to optionally set up their own sharing servers and configure access control mechanisms.

\subsection{Embedding into Web Pages}

\begin{figure}[tb]
  \centering
  \includegraphics[width=\linewidth]{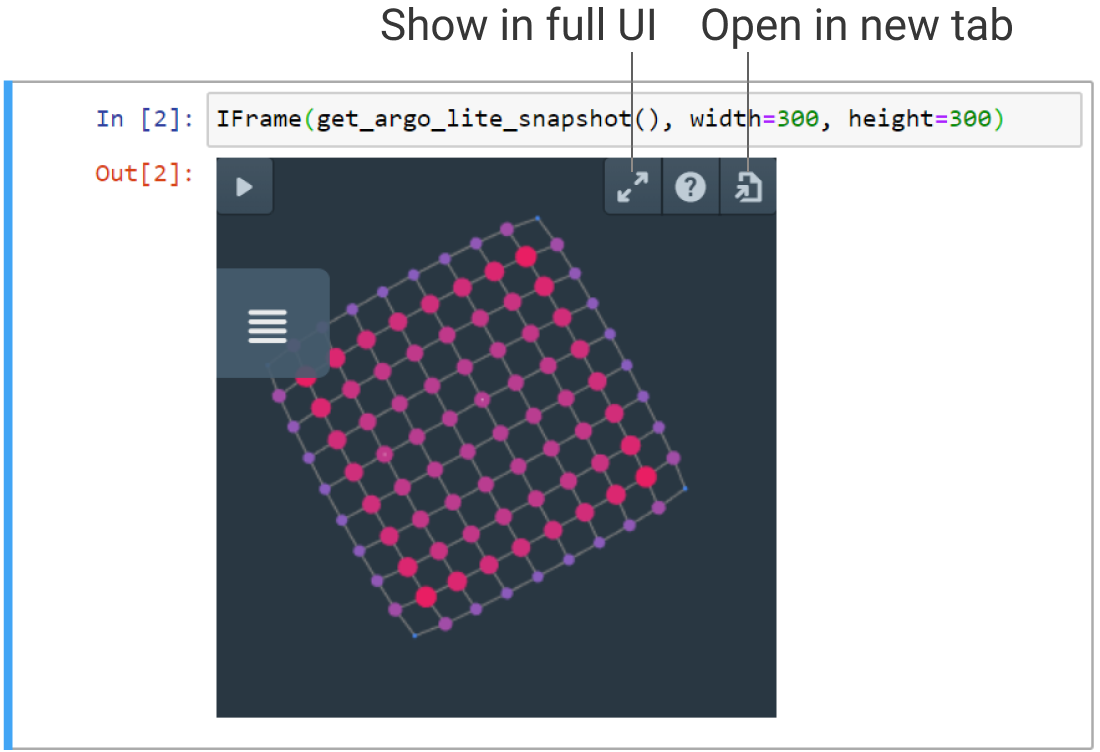}
  \caption{A Jupyter Notebook embedding an \argo{} widget. \argo{} makes publishing interactive graph visualization on the web much easier by offering an embeddable web widget based on iframe.}
  \Description{Figure showing Argo Lite's embedded web widget.}
  \label{fig:embed}
\end{figure}

When writing a blog post, sharing a graph dataset online, or working on an interactive notebook such as a Jupyter Notebook, it is useful to embed a graph in an interactive graph visualization viewer, such that the audience may easily engage in exploring the data themselves. 
\argo{} provides an embedded widget based on HTML iframe that allows users to easily embed any shared graph snapshots (\autoref{fig:embed}) onto HTML-based documents. Users can view and interact with the shared graphs within a webpage or (Jupyter) notebook. 
Users can expand the embedded widget into the full-featured \argo{} interface within the iframe, 
or open the graph snapshot in a new browser window or tab.

\section{System Design}
\label{sec:implementation}

\subsection{Web Client and User Interface}

The web UI of \argo{} is written completely in modern JavaScript (ES6+), using the mainstream React framework (\url{https://reactjs.org}) and the MobX state management library (\url{https://mobx.js.org}), making it easy to develop and maintain. On top of React, we use a combination of Blueprint.js and custom CSS to style and animate our UI, bringing a familiar interface that new users can quickly get used to and to ensure that our interface is reactive and mobile friendly.

\subsection{Efficient Rendering via WebGL}

Traditional browser rendering techniques using HTML canvas or SVG do not scale well with graphs larger than a few hundred nodes \cite{Carina}. We use WebGL, a powerful rendering solution that can take advantage of hardware acceleration and many other optimizations.

\argo{} has a dedicated WebGL renderer written in vanilla JavaScript with minimal dependencies. It interacts with WebGL through the Three.js library. The renderer provides a complete set of APIs that work with the rest of the React application. It can be easily isolated and used within another application.

\subsection{Sharing Server}

\argo{} provides and hosts a free sharing server on Heroku implemented using Node.js and the open source Strapi content management system. The free sharing server allows anyone to share their data and visualization to the public world.

For researchers who may be working on sensitive or proprietary data,
\argo{} also works wth their own preferred backend servers.
We have provided the documentation and source code for the users who need to
deploy their servers and implement custom access control policies.

\section{Availability and Utility}

\subsection{Open Source Code and License}
\argo{} is completely open source, with source code and documentation available at the Github repository \textcolor{linkColor}{\url{https://github.com/poloclub/argo-graph-lite}}. The code is licensed under the permissive MIT license, freely available to everyone. 

\subsection{Documentation}
We have provided extensive documentation for using, developing and customizing the software. Documentation and tutorials are available at the GitHub repository \textcolor{linkColor}{\url{https://github.com/poloclub/argo-graph-lite}}.

\subsection{Real-world Example Graph: Citation Graph of COVID-19 Research Publications} \label{sec:covid}
\argo{} comes with multiple example graphs available under the \textit{Graph} menu.
In addition to the classic example \textit{Les Misérables} character graph \cite{konect:knuth1993}, which visualizes the interactions between major characters in the Les Misérables novel, 
\argo{} also provides the real-world citation graph of recent COVID-19 publications derived from the CORD-19 Open Research Dataset \cite{Wang2020CORD19TC}. 
In this citation graph (\autoref{fig:teaser}), each node represents a published paper related to COVID-19, and each edge represents a citation relationship between two papers.
As our world is going through the COVID-19 health crisis, there is a surge of research publications on this matter, with researchers from diverse fields working to investigate it.
Since the papers included in this dataset come from many different fields of study, researchers might be interested in finding publications more relevant to their interests.
\argo{} is capable of hiding irrelevant papers.
Researchers are able to start exploration from highly cited papers in their specific fields of interest, and incrementally add neighbor nodes of these publications.
This helps researchers build subgraphs of publications tailored to their own  research interests.

\subsection{In-class Usage at Georgia Tech} \label{sec:class}
\argo{} may be used by people without prior experience or expertise in graph visualization.
It has been successfully used by more than 1,000 students in Georgia Tech's \textit{Data and Visual Analytics} class for learning graph visualizations. Students have learned to use force-directed layout and different sizing, coloring and labeling options to visualize their graphs. Additionally, students shared their graph visualizations as URLs using \argo{}'s link sharing functionality. Thus, we expect \argo{} to be a resource not only useful for experienced researchers, but also valuable for students learning about graph analysis and visualization techniques.

\section{Predicted Impact}

\subsection{CIKM Research Activities Enabled and Enhanced}

\argo{} has strong potential to help advance multiple CIKM research areas, as highlighted on \href{https://cikm2020.org/call-for-papers-full-and-short-research-papers}{CIKM's website}.
\argo{} is a 
modern
tool for \textit{data presentation}, particularly for visualizing and exploring graphs.
\argo{} provides an easy-to-use \textit{user interface for interactive analysis}, and could help users more easily explore knowledge graphs.
As demonstrated through the example citation graph of COVID-19 publications in Section \ref{sec:covid},
\argo{} may also help with \textit{data integration and aggregation}, assisting researchers and practitioners who develop and share knowledge graphs.

\subsection{Advancing Graph Visualization and Exploration}
Graph visualization has been a well-established field growing in importance, and popular tools such as Gephi \cite{bastian2009gephi} and Cytoscape \cite{shannon2003cytoscape} have already existed for many years. \argo{} seeks to advance graph visualization and exploration by bringing the power of graph visualization into a modern in-browser web app available on both desktop and mobile devices, equipped with both the classic visualization features as well as new sharing and collaboration features.
\argo{} addresses the challenges of existing tools in terms of availability and interactivity, and offers a modern easy-to-use experience.
It is a valuable addition not only to the graph visualization field, but also other research areas that benefit from graph visualizations.

\subsection{Timeline}
We expect \argo{} to be useful for a long time and remain updated as an evolving open source project. We will continuously maintain and improve \argo{} and will leverage GitHub's issue board to respond to user feedback. Meanwhile, \argo{} will also continue to be used in Georgia Tech's Data and Visual Analytics class, for use by over 2,000 students every year (1,000 per semester).

As more and more graph data are collected and made available, there's a growing interest to visualize, explore and understand graph data. We anticipate that more researchers and students will be using \argo{} as a convenient graph visualization and exploration tool in the coming years.

\section{Conclusion and Discussion}
\argo{} is a novel open-source in-browser interactive graph exploration and visualization tool. It enables researchers to incrementally explore graph data in browser and conveniently share their interactive visualizations via URLs and embedded widgets.

Since \argo{} is an in-browser tool, it does have limitations, since browser web apps have limited memory and cannot access file systems directly. In order for \argo{} to be useful for even larger graphs, such as those with millions of edges, we anticipate future development of backend web services responsible for storing and processing large graph files uploaded by users. These services may be paired with a dedicated frontend based on \argo{} for interactive visualization and exploration. It would also be useful to see \argo{} being used as a common frontend integrated into popular graph databases.

We believe that \argo{} is a valuable tool for both researchers and students alike, and will continue to maintain and improve this open-source project.

\bibliographystyle{ACM-Reference-Format}
\bibliography{main}

\end{document}